\documentclass[fleqn,usenatbib]{mnras}

\usepackage{newtxtext,newtxmath}

\usepackage[T1]{fontenc}

\DeclareRobustCommand{\VAN}[3]{#2}
\let\VANthebibliography\thebibliography
\def\thebibliography{\DeclareRobustCommand{\VAN}[3]{##3}\VANthebibliography}

\usepackage{graphicx}	
\usepackage{amsmath}	
\usepackage{multirow}
\usepackage{orcidlink}

\title[Redshift evolution of the MZR]{Inflow and outflow properties, not total gas fractions, drive the evolution of the mass-metallicity relation}

\author[L. Bassini et al.]{\parbox{\textwidth}{
Luigi Bassini\orcidlink{0000-0002-6864-7762},$^{1}$
Robert Feldmann\orcidlink{0000-0002-1109-1919},$^{1}$\thanks{E-mail: robert.feldmann@uzh.ch}
Jindra Gensior\orcidlink{0000-0001-6119-9883},$^{1}$
Claude-André Faucher-Giguère\orcidlink{0000-0002-4900-6628},$^{2}$
Elia Cenci\orcidlink{0000-0002-0766-1704},$^{1}$
Jorge Moreno\orcidlink{0000-0002-3430-3232},$^{3, 4}$\thanks{Flatiron IDEA Scholar}
Mauro Bernardini\orcidlink{0000-0002-2930-9509},$^{1}$
Lichen Liang\orcidlink{0000-0001-9422-0095}$^{5}$}
\\
$^{1}$
Department of Astrophysics, University of Zurich, Zurich CH-8057, Switzerland\\
$^{2}$
Department of Physics and Astronomy and CIERA, Northwestern University, 2145 Sheridan Road, Evanston, IL 60208, USA\\
$^{3}$
Department of Physics and Astronomy, Pomona College, Claremont, CA 91711, USA\\
$^{4}$
Center for Computational Astrophysics, Flatiron Institute, 162 Fifth Avenue, New York, NY 10010, USA\\
$^{5}$Canadian Institute for Theoretical Astrophysics, University of Toronto, 60 St. George Street, Toronto, ON M5S 3H8, Canada\\
}

\date{Accepted XXX. Received YYY; in original form ZZZ}

\pubyear{2015}

\begin{document}
\label{firstpage}
\pagerange{\pageref{firstpage}--\pageref{lastpage}}
\maketitle

\begin{abstract}
Observations show a tight correlation between the stellar mass of galaxies and their gas-phase metallicity (MZR). This relation evolves with redshift, with higher-redshift galaxies being characterized by lower metallicities. Understanding the physical origin of the slope and redshift evolution of the MZR may provide important insight into the physical processes underpinning it: star formation, feedback, and cosmological inflows. While theoretical models ascribe the shape of the MZR to the lower efficiency of galactic outflows in more massive galaxies, what drives its evolution remains an open question. In this letter, we analyze how the MZR evolves over $z=0-3$, combining results from the FIREbox cosmological volume simulation with analytical models. Contrary to a frequent assertion in the literature, we find that the evolution of the gas fraction does not contribute significantly to the redshift evolution of the MZR. Instead, we show that the latter is driven by the redshift-dependence of the inflow metallicity, outflow metallicity, and mass loading factor, whose relative importance depends on stellar mass.
These findings also suggest that the evolution of the MZR is not explained by galaxies moving along a fixed surface in the space spanned by stellar mass, gas phase metallicity, and star formation rate.
\end{abstract}



\section{Introduction}

The amount of metals within the interstellar medium (ISM) is set by the current and past star formation rate (SFR), the magnitude and chemical enrichment of galactic inflows from the circum-galactic medium (CGM), and the strength of galactic outflows that remove metals from the ISM (e.g., \citealt{2011PEEPLES}, \citealt{2012DAVE}, \citealt{2013LILLY}, \citealt{2017DEROSSI}, \citealt{2019TORREY}, and \citealt{2019MAIOLINO} for a recent review). It thus provides a critical benchmark for theoretical models of galaxy formation and evolution.

Observationally, the gas phase oxygen abundance $\rm (O/H)$ is tightly linked to the galaxy stellar mass ($M_{\star}$), with lower metallicities found in less massive galaxies (e.g., \citealt{1979LAEQUEX}, \citealt{2004TREMONTI}, \citealt{2006LEE}, \citealt{2008KEWLEY}, \citealt{2012BERG}, \citealt{2013ANDREWS}, \citealt{2019BLANC}, \citealt{2020CURTI}). Moreover at fixed $M_{\star}$, galaxies at higher redshift are characterized by lower gas metallicities (e.g., \citealt{2005SAVAGLIO}, \citealt{2006ERB}, \citealt{2008MAIOLINO}, \citealt{2009MANNUCCI}, \citealt{2014CULLEN}, \citealt{2014MAIER}, \citealt{2014STEIDEL}, \citealt{2014TRONCOSO}, \citealt{2016ONODERA}, \citealt{2021SANDERS}).

From a theoretical perspective, the stellar mass dependence of the mass-metallicity relation (MZR), as well as its redshift evolution are frequently studied either via analytical models (e.g., \citealt{2008FINLATOR}, \citealt{2011PEEPLES}, \citealt{2012DAVE}, \citealt{2013LILLY}, \citealt{2013DAYAL}, \citealt{2015FELDMANN_Z}), or with cosmological simulations and semi-analytical models (e.g., \citealt{2011DAVE}, \citealt{2016MA}, \citealt{2017DEROSSI}, \citealt{2019TORREY}, \citealt{2021FONTANOT}). The analytic models are constructed around the conservation of baryonic mass within galaxies, and they are generally able to describe both the shape and the redshift evolution of the MZR, although they resort to different physical interpretations. While there is general consensus that a more efficient expulsion of metals from lower mass galaxies sets the slope of the MZR
(although \citealt{2023BAKER} argue that it is a consequence of the stellar mass being proportional to the overall metals produced in the galaxy), what drives the evolution of the MZR is still debated. Specifically, some models find that this evolution is mainly driven by more enriched gas inflows at lower redshift (e.g., \citealt{2012DAVE}), while others relate the evolution to different SFRs (or, equivalently, gas masses) at fixed $M_{\star}$ at different redshifts (e.g., \citealt{2013LILLY}). The latter is consistent with the existence of a fundamental plane for metallicity (e.g., \citealt{2010MANNUCCI}). In this view, the MZR is a 2D projection of a 3D plane consisting of $M_{\star}-Z-$SFR (or $M_{\star}-Z-M_{\rm gas}$), and the redshift evolution of the MZR is a consequence of the redshift evolution of the average gas masses and SFRs in galaxies.

Similar results are also found with hydrodynamical simulations. Indeed, there is a general consensus on the role of feedback in setting the slope of the MZR. Specifically, \cite{2017DEROSSI} used different variations of the EAGLE  galaxy formation model, showing that at $M_{\star} \lesssim 10^{10} M_{\odot}$ the slope of the MZR is mainly set by stellar feedback, while feedback from active galactic nuclei (AGN) plays a major role at larger stellar masses. Similar results were also found by \cite{2011DAVE}. However, as for analytical models, no general consensus on the physical properties leading to the redshift evolution of the MZR has been reached. While in the EAGLE and IllustrisTNG models this evolution is attributed to evolving gas fractions or SFR (\citealt{2017DEROSSI}, \citealt{2019TORREY}), \cite{2011DAVE} argued that the main physical property driving the evolution is the metallicity of the inflowing material.

In this paper, we combine results from a state-of-the-art cosmological volume simulation (FIREbox, \citealt{2023FELDMANN}) with analytic models to study the physical mechanisms driving the redshift evolution of the MZR. By using a large set of galaxies from a cosmological volume, we are able to study galactic properties in a statistical manner. The physics model (FIRE-2, \citealt{2018HOPKINS}) employed in FIREbox is well suited to explore the gas-phase metallicity since it is able to resolve the ISM and produces galactic outflows self-consistently (\citealt{2015MURATOV}, \citealt{2017ALCAZAR}, \citealt{2017MURATOV}, \citealt{2021PANDYA}). Specifically, unlike most of other currently available full-box simulations where galactic winds are free parameters of the subgrid models, in FIRE galactic winds emerge from multi-channel stellar feedback implemented on the scale of star-forming regions. This implies that wind mass and metal loading factors emerge from the local injection of energy and momentum and are not prescribed or tuned. 
In the context of galactic metallicities, \citealt{2016MA}, using a set of zoom-in cosmological simulations showed that this model produces gas-phases metallicities that agree reasonably well with observations in the redshift range $0\leq z \leq 6$.

\section{Simulations}

In this letter, we study the properties of galaxies relevant to the MZR and its evolution drawn for the FIREbox cosmological volume (22.1 Mpc)$^3$ simulation (\citealt{2023FELDMANN}). The simulation is part of the Feedback In Realistic Environments (FIRE) project\footnote{\url{https://fire.northwestern.edu/}}, and it was run with the cosmological code GIZMO\footnote{\url{http://www.tapir.caltech.edu/~phopkins/Site/GIZMO.html}} (\citealt{2015HOPKINS}) using the Meshless Finite Mass hydro solver and the FIRE-2 physics (\citealt{2018HOPKINS}). Specifically, gas cooling and heating rates are computed for temperatures ranging from $10-10^9$ K, with the inclusion of heating and photoionization from a \cite{2009FAUCHER} UV Background. Stars form
from gas particles with a local efficiency of 100 percent per free-fall time if gas particles are: self-gravitating, Jeans unstable, and above a density threshold of $300\ \rm cm^{-3}$. The simulations implement different stellar feedback channels. Specifically: feedback from SN of type II and Ia, stellar winds from massive OB and evolved AGB stars, photoionization, photoelectric heating, and radiation pressure. In the simulation, we track 15 chemical species (H, He, C, N, O, Ne, Mg, Si, S, Ca, Fe, and four tracker species for r-process elements) and we include sub-grid
metal diffusion from unresolved turbulence (\citealt{2017SU}, \citealt{2018ESCALA}).

FIREbox is run at a mass resolution of $m_{\rm b} = 6.3\times 10^{4} M_{\odot}$  and $m_{\rm DM} = 3.3\times 10^{5} M_{\odot}$ for gas and dark matter particles respectively. Star particles form from gas particles and maintain the progenitor particle mass. The values of the softening lengths for star particles (DM particles) are $\epsilon_{\star}=12$ pc ($\epsilon_{\rm DM}$ = 80 pc). The softening length for gas particles is adaptive, with a fixed minimum value of 1.5 pc. The softening lengths are fixed in proper (comoving) units at $z<9$ ($z\geq 9$). In this letter, we make use of all central galaxies with a stellar mass $M_{\star} > 10^{8} M_{\odot}$, identified in the four redshift bins $z=0, 1, 2, 3$. Galaxies are identified with the AMIGA halo finder (AHF, \citealt{2004GILL}, \citealt{2009AKNOLLMANN}).

\section{Mass metallicity relation in FIREbox}

\begin{figure}
    \centering
    \includegraphics[width=1.0\linewidth]{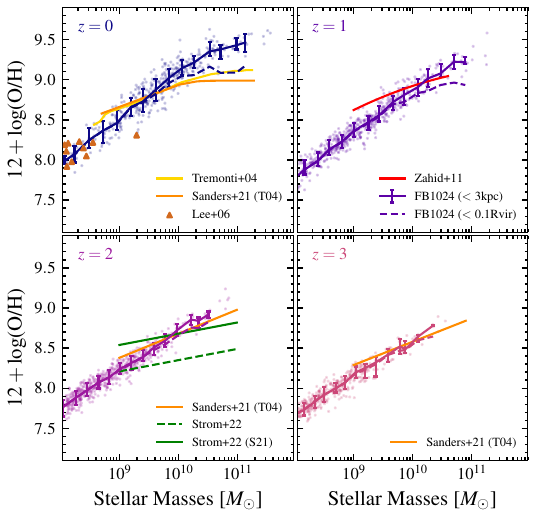}
    \caption{MZR in observations and simulations at $0 \leq z \leq 3$. In simulations, stellar masses are computed within $0.1\times R_{\rm vir}$. Metallicity is computed within a spherical aperture of $3$ kpc. Values for single galaxies are shown by lightly coloured circles. Solid lines show the median of the distribution, with error bars encompassing the $16^{\rm th}-84^{\rm th}$ percentile region. As a reference to show the dependence of the results on the region within which the metallicity is computed, we also show the results within $0.1 \times R_{\rm vir}$ as coloured dashed lines. For observations, we show the results with coloured triangles and lines, as described in the legend. The data from \protect\cite{2021SANDERS} are rescaled to match the calibration of \protect\cite{2004TREMONTI} (see text for more details). We show the MZR measured by \protect\cite{2022STROM} both with their original normalization (dashed green line), and by normalizing them to match the \protect\cite{2021SANDERS} results at $M_{\star} = 10^{10} M_{\odot}$. Simulation results agree reasonably well with observations at all redshifts apart from the massive end at $z=0$ (when using the 3 kpc apertures that roughly match the aperture used in the sample of \protect\citealt{2004TREMONTI}), and the slope of the MZR at $z=2$ which is steeper in simulations.}
    \label{fig:MZR_observations}
\end{figure}

\begin{figure*}
    \centering
    \includegraphics[width=1.0\linewidth]{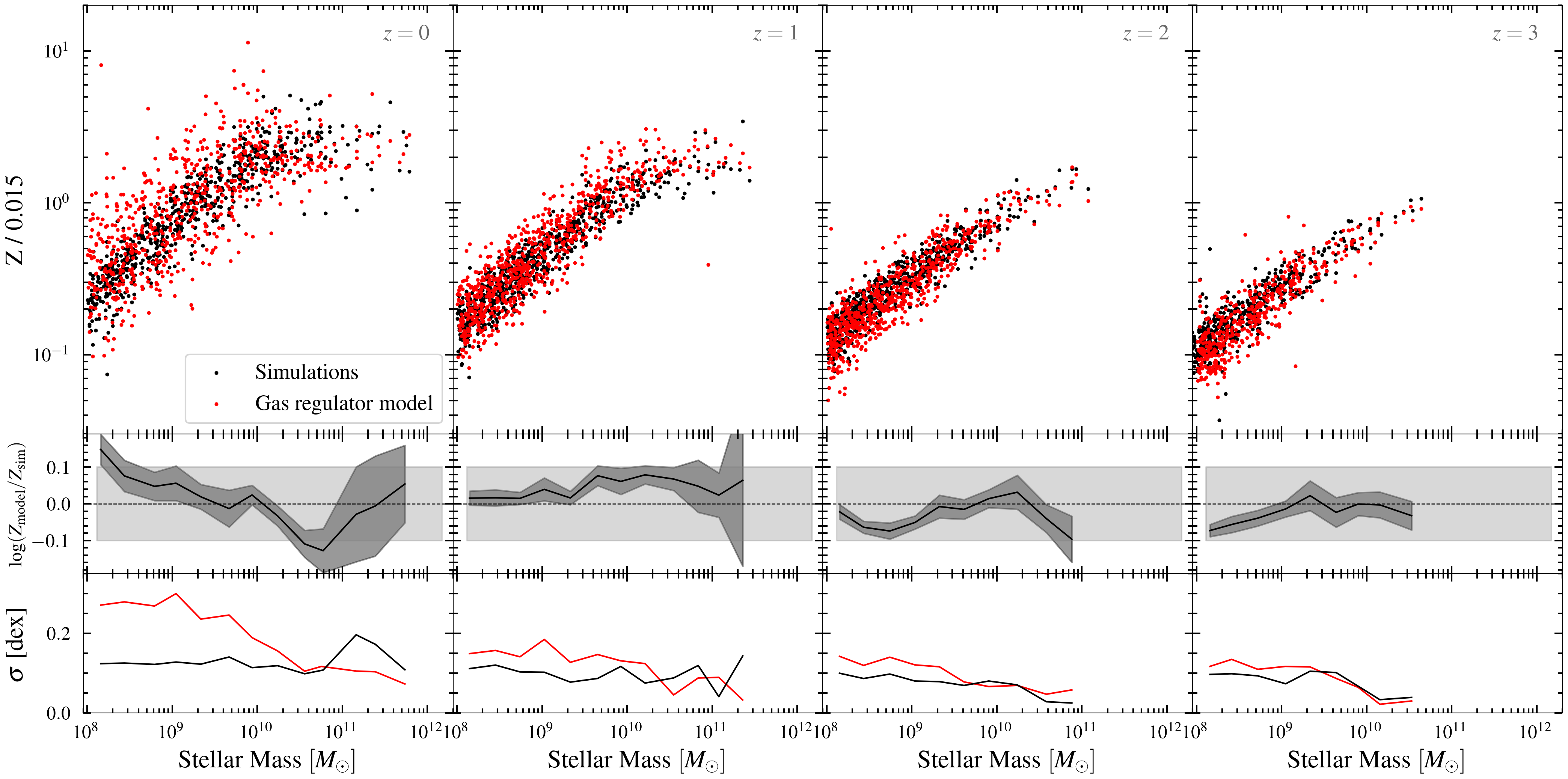}
    \caption{Comparison between the metallicity of FIREbox galaxies (black points) and the prediction from the analytical model (red points) given in Eq.~\ref{eq:equilibrium_model} at $z=0, 1, 2, 3$ (top row, from left to right). All the terms entering Eq.~\ref{eq:equilibrium_model} are directly computed from the simulations as described in Appendix~\ref{Appendix:eq_model}. The middle panels show the logarithm (in base 10) of the ratio between the median values of the metallicity from the analytic model and the metallicity directly measured from the simulation, as solid black line. The darker shaded region around the curve shows the $1-\sigma$ interval obtained from bootstrapping, while the dotted black line and the light grey shaded region around it indicate exact agreement between the model and the simulation and the $0.1$ dex difference interval respectively. The lower panels show the scatter around the median values, defined as half the difference between the 84$^{\rm th}$ and 16$^{\rm th}$ percentiles. The analytical model well describes the metallicity of FIREbox galaxies, with median values in agreement within $0.1$ dex.}
    \label{fig:gas_regulator_model}
\end{figure*}

\begin{figure}
    \centering
    \includegraphics[width=0.9\linewidth]{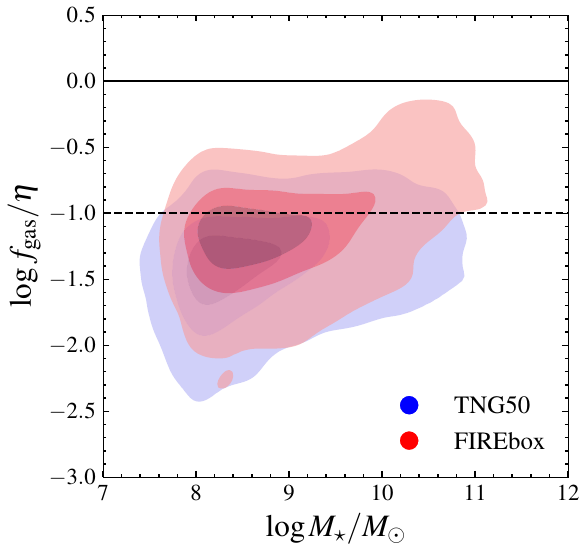}
    \caption{Ratio between gas fraction, $f_{\rm gas}$, and mass loading factor, $\eta$, as a function of stellar mass for FIREbox (red) and IllustrisTNG (blue) for combined redshifts $z=0-3$. The ratio is typically lower than $0.1$, implying that the variation in the gas fraction required to explain the redshift evolution of the MZR is much larger than what state-of-the-art cosmological simulations predict (see Eqs.~\ref{eq:equilibrium_model}, ~\ref{eq:r_gas_fraction} and the text for further details). }
    \label{fig:fgas_vs_eta.pdf}
\end{figure}

\begin{figure*}
    \centering
    \includegraphics[width=1.0\linewidth]{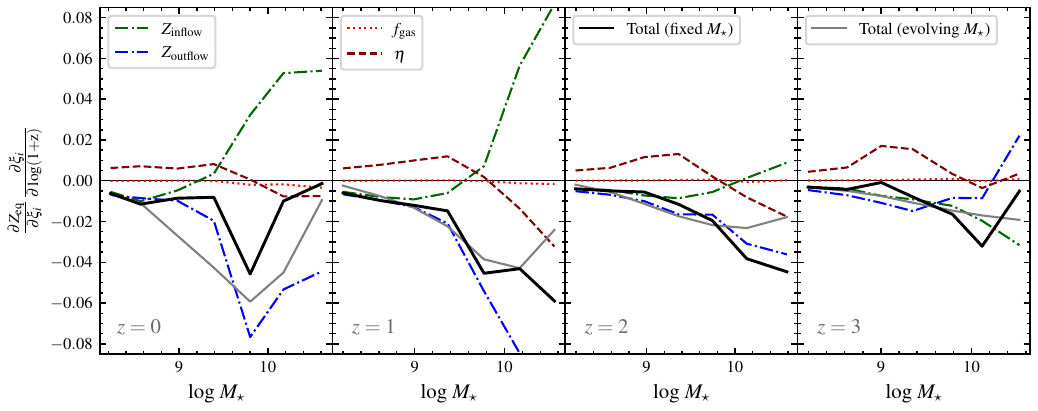}
    \caption{Redshift evolution of the mass-metallicity relation as driven by different physical quantities and at different redshifts. The different colors with different line styles show the four terms in Eq.~\ref{eq:analytical_espression}. Specifically: inflow metallicity, $Z_{\rm inflow}$ (green, dashdot), outflow metallicity, $Z_{\rm outflow}$ (blue, densely dashdotted), gas fraction (red, dotted), and mass loading factor, $\eta$ (maroon, dashed). We show in black the sum of all the terms. In grey, we also show the evolution of gas phase metallicity as driven by the increment of the stellar mass with time (second term of Eq.~\ref{eq:analytical_espression}, see text for more details). The four different panels refer to $z=0, 1, 2, 3$, from left to right. Negative values imply that the specific factor induces a decrement in the normalization of the MZR going to higher redshifts. From the plot, we can see that the decreasing normalization of the MZR depends on different physical properties in different mass regimes. At $M_{\star} \lesssim 10^{10}\ M_{\star}$, the main driver is the metallicity of the outflows, with a comparable contribution from the metallicity of the inflows. At larger stellar masses, the contribution from inflow metallicity changes signs, partially compensating for the evolution driven by the outflow metallicity. At all redshifts, the contribution to the evolution of galaxy metallicity from the evolving $M_{\star}$ is comparable to the evolution driven by the $z$ dependence of $\xi_i$. }
    \label{fig:redshift_evolution}
\end{figure*}

In Fig.~\ref{fig:MZR_observations} we show the MZR in FIREbox in comparison with observational data in the redshift range $0\leq z \leq3$. In the simulations, stellar masses are computed within $0.1\times R_{\rm vir}$ (where the virial radius is computed following the virial overdensity definition of \citealt{1998BRYAN}).  Metallicities are computed as the average gas-phase Oxygen to Hydrogen abundance ratios within two different apertures: 3 kpc (solid lines; this roughly matches the aperture used in the sample of \citealt{2004TREMONTI}) and $0.1\times R_{\rm vir}$ (dashed lines; this roughly matches galaxy sizes). Results from simulations are shifted downward by 0.12 dex in order to account for oxygen depletion inside HII regions (\citealt{2010PEIMBERT}, \citealt{2022FELDMANN}). Regarding the observational samples, we take the results of \cite{2004TREMONTI}, \cite{2006LEE}, and \cite{2011ZAHID} at face value as they employ similar metallicity calibrations (\citealt{2008KEWLEY}). We rescale the $z=0$ results from \cite{2021SANDERS} in order to match \cite{2004TREMONTI} MZR at $M_{\star}=10^{10} M_{\odot}$. We then apply the same normalization factor to \cite{2021SANDERS} data at $z>0$. Finally, we plot the data of \cite{2022STROM} at $z\sim 2.3$ both at face value (dashed green line) and matching the normalization of \cite{2021SANDERS} at $M_{\star}=10^{10} M_{\odot}$ (solid green line).

Figure~\ref{fig:MZR_observations} shows that data from FIREbox agree reasonably well with observations in the redshift range covered, given the substantial systematic uncertainties in observational metallicity measurements (e.g., \citealt{2019MAIOLINO}). The only exceptions are represented by the excess in metals in massive FIREbox galaxies ($M_{\star}>10^{10}\ M_{\odot}$) at $z=0$, and by the slope of the relation at $z\sim 2$. The former is likely related to the absence of an AGN feedback model in FIREbox. For example, numerical experiments run with the EAGLE model have shown that the slope of the MZR at $M_{\star}\gtrsim 10^{10} \ M_{\odot}$ is mostly set by AGN feedback (\citealt{2017DEROSSI}). Regarding the latter, the two relations from \cite{2021SANDERS} and \cite{2022STROM} at $z\sim 2$ represent the range of slopes reported in the literature for the MZR at high redshift. As discussed in \cite{2022STROM}, the slope of the MZR is sensitive to both the choice of the calibration and the galaxy sample. However, while uncertainties in the observed MZR remain large, FIREbox MZR is shallower than most observed MZR. Future investigations will be needed to pinpoint the reason behind this difference. 

\section{The Equilibrium metallicity in analytic models}

Having assessed that FIREbox produces gas-phase metallicities that are in approximate agreement with observational data, we now investigate whether simple analytical models accurately describe the properties of the simulated galaxies. These models are based on baryonic mass conservation within galaxies. Specifically, we will use the models described in \cite{2013LILLY} and \cite{2015FELDMANN_Z}, as they allow all parameters to vary, including inflow and outflow metallicities. Assuming that metals are instantaneously recycled and that the mass outflow rate is directly proportional to the SFR, the gas phase metallicity can be expressed as (e.g., \citealt{2015FELDMANN_Z}):
\begin{equation}\label{eq:bathtub_model}
    Z = \frac{y(1-R)r - \dot{Z}t_{\rm dep}}{1 - r_{Z}^{\rm in} + (r_Z^{\rm out} - 1)r\eta},
\end{equation}
 where
\begin{equation}\label{eq:r_factor_eta}
    r \equiv \frac{\rm SFR}{\dot{M}_{\rm gas, in}} = \frac{1}{1- R + \eta + t_{\rm dep} \left[ (1-R)\rm sSFR + \frac{\rm d \ ln \ SFR}{dt} + \frac{\rm d \ ln \ t_{\rm dep}}{dt}  \right]}, 
\end{equation}
(see also \citealt{2013LILLY}). 

In Eq.~\ref{eq:bathtub_model}, $y$ is the metal yield, $R$ is the return fraction of gaseous material from the formed stars in the instantaneous recycling approximation, $t_{\rm dep}$ is the gas depletion time defined as $t_{\rm dep}=M_{\rm gas} / \rm SFR$, $r_Z^{\rm in}=Z_{\rm inflow}/Z_{\rm ISM}$ ($r_Z^{\rm out}=Z_{\rm outflow}/Z_{\rm ISM}$) is the metallicity of the inflows (outflows) with respect to the metallicity of the ISM, $\dot{M}_{\rm gas, in}$ and $\dot{M}_{\rm gas, out}$ are inflow and outflow rate respectively, sSFR is the specific SFR (SFR$/M_{\star}$), and $\eta$ is the mass loading factor defined as $\eta = \dot{M}_{\rm gas, out} / \rm SFR$. Importantly, $M_{\rm gas}$ is the total gas mass including molecular, atomic, and ionized components. While different phases are generally correlated with each other in the local Universe (e.g., \citealt{2022SAINTOGE}), their redshift evolution might be considerably different. Indeed, observations show that the evolution of the cosmic mass fraction of atomic hydrogen is much weaker than the molecular one (e.g., \citealt{2020PEROUX}, \citealt{2020WALTER}).

If $\dot{Z}$ is much shorter than the depletion time (i.e., the time scale over which the metallicity evolves is much longer than the depletion time), then the second term in the numerator of Eq.~\ref{eq:bathtub_model} is negligible and it is possible to express the equilibrium metallicity as:

\begin{equation}\label{eq:equilibrium_model}
    Z_{\rm eq} = \frac{y(1-R)}{1 - r_{Z}^{\rm in} + (r_Z^{\rm out} - 1)r\eta}r.
\end{equation}

Moreover, if the time scale over which galaxy-integrated properties vary is long (i.e., galaxies are in equilibrium), the time derivatives in Eq.~\ref{eq:r_factor_eta} can be dropped and $r$ can be written as:

\begin{equation}\label{eq:r_gas_fraction}
    r = \frac{1}{1- R + \eta + (1-R)f_{\rm gas}},
\end{equation}
where $f_{\rm gas}$ is the gas fraction, $f_{\rm gas} = M_{\rm gas} / M_{\star}$, $M_{\rm gas}$ being the total gas mass. This gas fraction is not directly comparable to the one reported in observations of medium-to-high redshift galaxies since the latter measures primarily the cold, largely molecular ISM (\citealt{2020TACCONI}). We confirm that the approximation of Eq.~\ref{eq:r_factor_eta} given by Eq.~\ref{eq:r_gas_fraction} is indeed valid for FIREbox galaxies, see Appendix~\ref{Appendix:eq_model}.


Eqs.~\ref{eq:equilibrium_model}, \ref{eq:r_gas_fraction} imply that the MZR can evolve with redshift as a consequence of $(i)$ redshift-dependent inflow/outflow metallicities (Eq.~\ref{eq:equilibrium_model}), $(ii)$ redshift-dependent gas fractions (Eq.~\ref{eq:r_gas_fraction}), and $(iii)$ redshift-dependent values of the mass loading factor (Eqs.~\ref{eq:equilibrium_model}, ~\ref{eq:r_gas_fraction}). The main goal of this letter is to investigate which of these mechanisms is the main driver of the redshift evolution of the MZR in FIREbox. 

\section{The analytical model applied to FIREbox galaxies}

The first step is to study whether Eq.~\ref{eq:equilibrium_model} accurately describes the metallicity of FIREbox galaxies. In Fig.~\ref{fig:gas_regulator_model} we show the comparison between the gas-phase metallicity of FIREbox (black points) and the metallicity as predicted by Eq.~\ref{eq:equilibrium_model} (red points). We describe how to compute all the terms entering Eq.~\ref{eq:equilibrium_model} in Appendix~\ref{Appendix:eq_model}. In short, all the quantities are directly computed from the simulation, without the introduction of any ad hoc scaling factors. Furthermore, all quantities are averaged over one depletion time\footnote{The depletion time is computed as the average depletion time of all FIREbox galaxies at a given redshift to smooth out the large variability introduced by the SFR.}. This is crucial as the analytical models consider galaxies to be in equilibrium, and the metallicity approaches its equilibrium value on a depletion timescale (\citealt{2013LILLY}). Reducing the averaging time results in an increase in the scatter of the predicted metallicities. 

Fig.~\ref{fig:gas_regulator_model} shows that the results from the analytical model match well the true metallicities measured from the simulations, with the median of the two distributions being in agreement within $0.1$ dex (as shown in the middle panels). This implies that the metallicity of FIREbox galaxies is near equilibrium, and justifies the assumption made to derive Eq.~\ref{eq:equilibrium_model} from Eq.~\ref{eq:bathtub_model} (i.e., neglecting the $\dot{Z}$ term). Furthermore, the bottom panels also show that the scatter of the two distributions is comparable at $z \gtrsim 1$. At $z=0$, the scatter relative to the model at $M_{\star} < 10^{10}M_{\odot}$ is a factor of $2-2.5$ larger than that of the simulated galaxies. We speculate that this effect is driven by a more rapid evolution of $Z$ at lower redshift in the simulation (see, e.g., the discussion of Fig.~\ref{fig:redshift_evolution}), implying a non-negligible contribution of $\dot{Z}$ in Eq.~\ref{eq:bathtub_model}. However, further investigation outside the scope of this letter is required to fully understand the large scatter at $z=0$. Despite the differences at $z=0$, the agreement shown in Fig.~\ref{fig:gas_regulator_model} is remarkable, considering the necessary simplifying assumptions needed in the analytical model (such as the instantaneous recycling approximation and a direct proportionality between SFR and mass outflow rate). This comparison demonstrates that analytical models correctly describe the average metallicity of simulated galaxies over a broad range of stellar mass and redshift.

\section{What drives the evolution of the MZR in FIREbox}

Given the success of the analytic model in reproducing the results of the cosmological simulation, we will now use the model to explore the drivers of the MZR evolution, in particular the role of the gas fraction. A first qualitative assessment can be made by considering the following simplified scenario with pristine inflowing material ($r^{\rm in}_Z=0$) and outflows with the same metallicity as the ISM ($r^{\rm out}_Z=1$). Under these assumptions, $Z_{\rm eq}\propto{}r$ according to Eq.~\ref{eq:equilibrium_model}, while $r$ depends significantly on $f_{\rm gas}$ only if $f_{\rm gas}$ is at least of the same order of magnitude as $\eta$, see Eq.~\ref{eq:r_gas_fraction}. 

According to Fig.~\ref{fig:fgas_vs_eta.pdf}, based on data from FIREbox as well as TNG50 \citep{2018PILLEPICH, 2019PILLEPICH}, the ratio between $f_{\rm gas}$ and $\eta$ is typically 0.1 or lower. Consequently, changing $r$ by a factor of 2 to match the observed evolution of the MZR between $z=0$ and $z=3$ \citep{2021SANDERS}, would require changing $f_{\rm gas}$ by a factor of 10 or more over the same redshift range. The actual change in gas fraction is, however, at most $\sim{}0.3$ dex in both FIREbox and TNG50, see Appendix~\ref{appendix:gas_fractions}. Therefore, $f_{\rm gas}$ is expected to play a negligible role in the redshift evolution of the MZR.


To strengthen this statement we also use a more quantitative approach. First, we note that following Eq.~\ref{eq:equilibrium_model} and Eq.~\ref{eq:r_gas_fraction}, the equilibrium metallicity depends upon four independent variables: $r_Z^{\rm in}$, $r_Z^{\rm out}$, $\eta$, and $f_{\rm gas}$. We will refer to these four variables as $\xi_i$, with $i$ running from $1$ to $4$. We find that the redshift dependence of y and R does not contribute at a significant level to the evolution of the metallicity, allowing us to ignore it in our further analysis.

For individual galaxies, $Z_{\rm eq} = Z_{\rm eq}(z, M_{\star}(z)) = Z_{\rm eq}(\xi_i(z, M_{\star}(z)))$, where the dependencies on redshift and stellar mass arise as possibly all four parameters depend on $z$ and $M_{\star}$. We also include a dependence of redshift on stellar mass, since the latter is allowed to increase with time. Hence, the redshift evolution of the equilibrium metallicity  can be written as:

\begin{equation}\label{eq:analytical_espression}
\begin{split}
    \frac{d Z_{\rm eq}}{d \log(1+z)} = \sum_{i=1}^{i\leq 4} \left( \frac{\partial Z_{\rm eq}}{\partial \xi_i} \left.\frac{\rm \partial \xi_i}{\rm \partial \log(1+z)}\right|_{M_{\star}} \right. + \\
    \left. \frac{\partial Z_{\rm eq}}{\partial \xi_i} \left. \frac{\rm \partial \xi_i}{\rm \partial M_{\star}}\right|_{z} \frac{d M_{\star}}{d \log(1+z)} \right).
\end{split}
\end{equation}

The first term in the parenthesis describes how the MZR evolves with redshift as a consequence of the evolution of $\xi_i$ at fixed stellar mass. The second term describes how galaxies evolve on the MZR as a consequence of the increase of their stellar mass with decreasing redshift.

The factors $\partial Z_{\rm eq} / \partial \xi_i$ can be computed analytically from Eqs.~\ref{eq:equilibrium_model}, ~\ref{eq:r_gas_fraction}. Given the analytical expression, the value of the derivative of $\xi_i$ at fixed redshift is then computed considering the median value of each independent variable $\xi_i$ in different mass bins taken directly from the simulation. Specifically, we compute the median values of $\xi_i$ in different mass bins at $z=0, 1, 2, 3$. Then, for each mass bin, we fit the data to $\log \xi_i = \alpha + \beta \times z + \gamma \times z^2$. The results of the fit, which are shown in Appendix~\ref{Appendix:z_fit}, are finally used to compute the derivative of $\xi_i$ with respect to redshift. A similar procedure is used to derive the terms in the second factor of Eq.~\ref{eq:analytical_espression} (in this case fitting $\xi_i$ as a function of $M_{\star}$ at fixed $z$, see Appendix~\ref{Appendix:z_fit}).

The results of this analysis are presented in Fig.~\ref{fig:redshift_evolution}. Specifically, we plot with coloured lines the four $(\partial Z_{\rm eq} / \partial \xi_i)(\partial \xi_i / \partial \log (1+z))$ terms. The black line shows the sum of these four terms, while the grey line shows $\sum_i (\partial Z_{\rm eq} / \partial \xi_i)(\partial \xi_i / \partial M_{\star})(dM_{\star}/d\log(1+z))$. Firstly, the results show that the changes in the gas fraction of galaxies (shown in red), do not play a major role in driving the redshift evolution of the MZR in FIREbox at any stellar mass and redshift analyzed in this letter. This is a fundamental difference with respect to other studies based on hydrodynamical cosmological simulations, where the redshift evolution is largely ascribed to $f_{\rm gas}$ (e.g., \citealt{2017DEROSSI}, \citealt{2019TORREY}). 

Instead, the main driver of the evolution in FIREbox is a combination of the metallicity of the outflows, the metallicity of the inflows, and the mass loading factor. Specifically, at  $M_{\star} \lesssim 3 \times 10^{9} M_{\odot}$, the main driver of the evolution is the metal content of inflows and outflows. Indeed, inflows (outflows) are more (less) metal-enriched with respect to the average ISM metallicity at lower redshift. The trend with outflow being more metal enriched with respect to the ISM at higher redshift is in line with previous FIRE results (\citealt{2017MURATOV}, \citealt{2021PANDYA}). Specifically, \cite{2021PANDYA} found that neither the mass loading factor nor the metal loading factor are redshift dependent (see their Fig.~5). This implies that the outflow metallicity is not strongly dependent on redshift (while the metallicity of the ISM is, as the MZR evolves with redshift).

At larger stellar masses, the contribution from inflow metallicity changes signs, implying that for massive galaxies inflows are more metal enriched (compared to ISM metallicity) at high redshift. This difference and transition at stellar masses $M_{\star} \lesssim 3 \times 10^{9} M_{\odot}$, can be interpreted in terms of gas recycling. \cite{2017ALCAZAR}, with a particle tracking analysis applied to zoom-in simulations run with the FIRE model, showed that the fraction of gas inflows coming from recycled gas decreases as stellar mass increases (see, e.g., their Fig.~6). This implies that at low stellar masses, most of the gas accreted through inflows will be pre-enriched at a metallicity comparable to the metallicity of the ISM. By contrast, in more massive galaxies, most of the inflowing gas will be pristine, thus effectively lowering the metallicity of the ISM. 

Finally, the contribution from galactic outflows becomes more relevant at $M_{\star} \gtrsim 3\times 10^9\ \rm M_{\odot}$. Since in Eq.~\ref{eq:equilibrium_model} the mass loading factor appears only at the denominator, the larger mass loading factor at higher redshifts in massive galaxies directly leads to this result.
 
In Fig.~\ref{fig:redshift_evolution}, we show in grey the contribution to the evolution of the metallicity due to the evolution of the stellar mass (the second term of Eq.~\ref{eq:analytical_espression}). From the analysis, we see that this contribution is comparable to the evolution driven by the redshift evolution of $\xi_i$. This implies that the metallicity evolution of galaxies is driven by both the evolution of its stellar mass, and the redshift dependence of inflow metallicities, outflow metallicities, and mass loading factors.

\section{Conclusions}

In this letter, we used the FIREbox cosmological simulation to study which physical quantities drive the redshift evolution of the MZR since cosmic noon within the FIRE-2 model for galaxy evolution. We have shown that FIREbox reproduces the mass-metallicity relation (MZR) reasonably well over the redshift range $0 \leq z \leq 3$ (see Fig.~\ref{fig:MZR_observations}), with the only tension represented by massive galaxies at $z=0$ and the slope of the relation at $z=2$. Moreover, we showed that the analytical model described in \cite{2013LILLY} and \cite{2015FELDMANN_Z} applied to FIREbox galaxies at redshifts $z=0, 1, 2, 3$ well reproduces the metallicity of simulated galaxies (see Fig.~\ref{fig:gas_regulator_model}). Given the values of the mass loading factors and gas fractions measured in cosmological simulations (see Fig.~\ref{fig:fgas_vs_eta.pdf}), we estimate that gas fractions need to increase by a factor of $10$ from $z=0$ to $z=3$ to explain the redshift evolution of the MZR. However, we find that gas fractions evolve at most by $0.3$ dex. In order to accurately interpret these findings, one must take into account that the gas fraction examined in this study is derived from the total gas mass, the redshift evolution of which may vary from that of individual gas phases, such as molecular gas. Finally, we used the analytical expression of the analytical model to determine which physical properties among the mass loading factor, $\eta$, the inflow and outflow metallicities (parametrized by $r_Z^{\rm in}$ and $r_Z^{\rm out}$ respectively) and the gas fraction $f_{\rm gas}$ represent the main driver of the redshift evolution of the MZR. The results show that, unlike commonly assumed, the gas fraction plays a negligible role in driving the redshift evolution of the MZR. Instead, in FIREbox the redshift evolution is mostly driven by redshift-dependent outflow metallicities, inflow metallicities, and mass loading factors, whose relative importance depends on galactic mass. 

The results shown in this paper imply that the redshift evolution of the MZR is the consequence of the redshift evolution of $r_Z^{\rm in}$, $r_Z^{\rm out}$, and $\eta$. This is fundamentally distinct from the commonly held view that the redshift evolution of the MZR is a manifestation of a fundamental plane with $M_{\star}-Z-\rm SFR$ (or $M_{\star}-Z-\rm f_{\rm gas}$) (e.g., \citealt{2010MANNUCCI}). We plan to investigate the link between our findings and the observational evidence for a fundamental plane in FIREbox in future work.

\section*{Acknowledgements}

LB thanks L. Boco for helpful discussions. LB, RF, EC, JG acknowledge financial support from the Swiss National Science Foundation (grant no PP00P2\_194814). RF, EC, MB acknowledge financial support from the Swiss National Science Foundation (grant no 200021\_188552). JG gratefully acknowledges financial support from the Swiss National Science Foundation (grant no CRSII5\_193826). CAFG was supported by NSF through grants AST-2108230  and CAREER award AST-1652522; by NASA through grants 17-ATP17-0067 and 21-ATP21-0036; by STScI through grant HST-GO-16730.016-A; and by CXO through grant TM2-23005X. JM is funded by the Hirsch Foundation. This work was supported in part by a grant from the Swiss National
Supercomputing Centre (CSCS) under project IDs s697 and s698. We
acknowledge access to Piz Daint at the Swiss National Supercomputing Centre, Switzerland under the University of Zurich’s share with
the project ID uzh18.  This work made use of infrastructure services provided by S3IT (\url{www.s3it.uzh.ch}), the Service and Support for Science IT team at the University of Zurich. All plots were created with the Matplotlib library for visualization with Python (\citealt{2007HUNTER}).

\section*{Data Availability}

 The data supporting the plots within this article are available on reasonable request to the corresponding author. A public version of the GIZMO code is available at \url{http://www.tapir.caltech.edu/~phopkins/Site/GIZMO.html}. FIRE-2 simulations are publicly available (\citealt{2022WETZEL}) at \url{http://flathub.flatironinstitute.org/fire}. Additional data, including initial conditions and derived data products, are available at \url{https://fire.northwestern.edu/data/}.



\bibliographystyle{mnras}
\bibliography{bibliography} 




\appendix

\section{The equilibrium model}\label{Appendix:eq_model}

\begin{figure}
    \centering
    \includegraphics{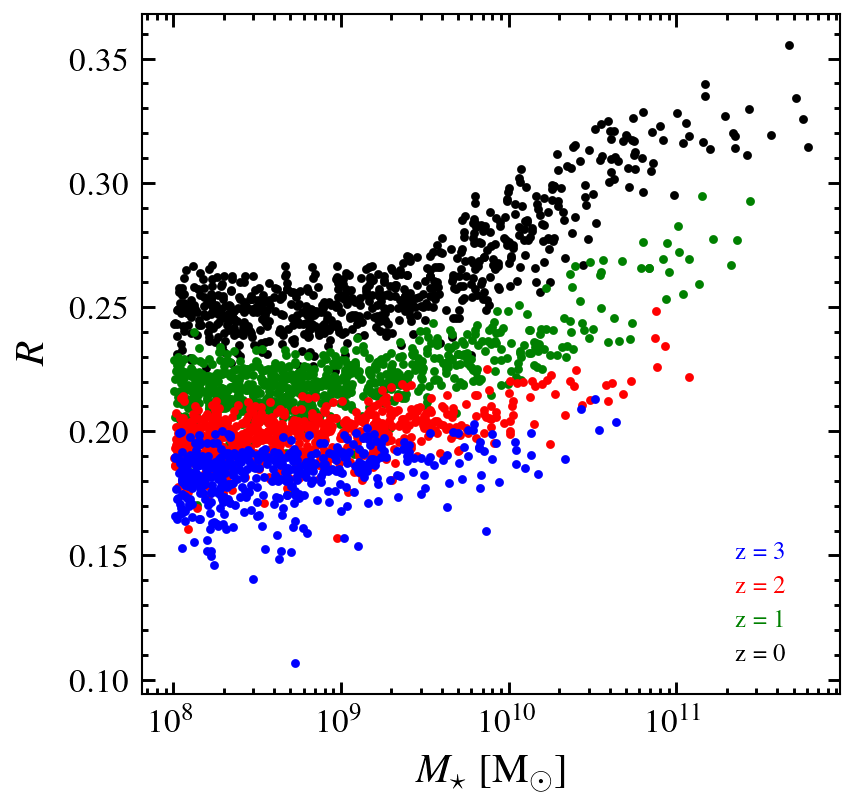}
    \caption{Returning fraction, R, of FIREbox galaxies at $z=0,1,2,3$. For each galaxy, R is computed as the mass-weighted average of the returning fraction of single star particles within $0.1\times R_{\rm vir}$.}
    \label{fig:R_Factor_SG}
\end{figure}

In this Appendix we describe how all the terms of Eq.~\ref{eq:equilibrium_model} and Eq.~\ref{eq:r_gas_fraction} are computed. Specifically:

\begin{itemize}

    \item $y$: effective yields are computed at each redshift by summing up all the metals in the cosmological volume (both in gas and stars), and normalizing by the total stellar mass. Using this definition we find $y=0.0375, 0.0327, 0.0291, 0.0289$ at $z=0,1,2,3$ respectively. 

    \item $R$: the return fraction is computed for each galaxy at all redshift bins considered. Specifically, we compute the value of $R$ for each star particle within a spherical region of $0.1\times R_{\rm vir}$. The value of $R$ for the galaxy is then obtained by computing the mass-weighted average of $R$ in the same spherical region. Values of $R$ for all FIREbox galaxies at $z=0, 1, 2, 3$ are shown in Fig~\ref{fig:R_Factor_SG}. 

    \item $\dot{M}_{\rm gas, in}$: To compute the mass inflow rate, we follow a particle tracing procedure. Specifically, to compute the inflow rate at a generic snapshot $s_i$, we consider all the gas particles that are within $0.1\times R_{\rm vir}$ at the snapshot $s_i$, and were outside the same spherical region at the snapshot $s_{i-1}$. This will return the overall gas mass within the gas inflows. To obtain the inflow rate, the gas mass is divided by the time separation between the two snapshots ($\sim 11$ Myr). 

    \item $\dot{M}_{\rm gas, out}$: here we follow a similar procedure as for $\dot{M}_{\rm gas, in}$. In this case, we consider all gas particles that are outside $0.1 \times R_{\rm vir}$ at the snapshot $s_{i}$ and were inside the same spherical region at the snapshot $s_{i-1}$. The mass outflow rate is then computed by dividing the gas mass within the outflows and the time separation between the two snapshots.

    \item $Z_{\rm in}$ and $Z_{\rm out}$: metallicities of the inflows and of the outflows are computed by dividing the total amount of metals carried by gas particles defined as inflowing or outflowing (see previous points) by the total gas mass in inflows and outflows.

    \item $M_{\rm gas}$ and $M_{\star}$: at each redshift, gas and stellar masses are computed considering gas and star particles within $0.1 \times R_{\rm vir}$.

\end{itemize}

Finally, all quantities are averaged over one depletion time. Since the SFR is highly variable, we use the average depletion time of all FIREbox galaxies at each redshift. This time average is used since it represents the time scale over which the equilibrium metallicity is reached (see Appendix of \citealt{2013LILLY}).

In Fig.~\ref{fig:r_factor} we also show that the approximation made to re-write Eq.~\ref{eq:r_factor_eta} into Eq.~\ref{eq:r_gas_fraction} is justified. Indeed, the two definitions follow the 1:1 line. The only difference is at large values of $r$, where the flattening is related to the saturation of the mass loading factor, $\eta$, that drops to values $\ll 1$.

\begin{figure}
    \centering
    \includegraphics[width=1.0\linewidth]{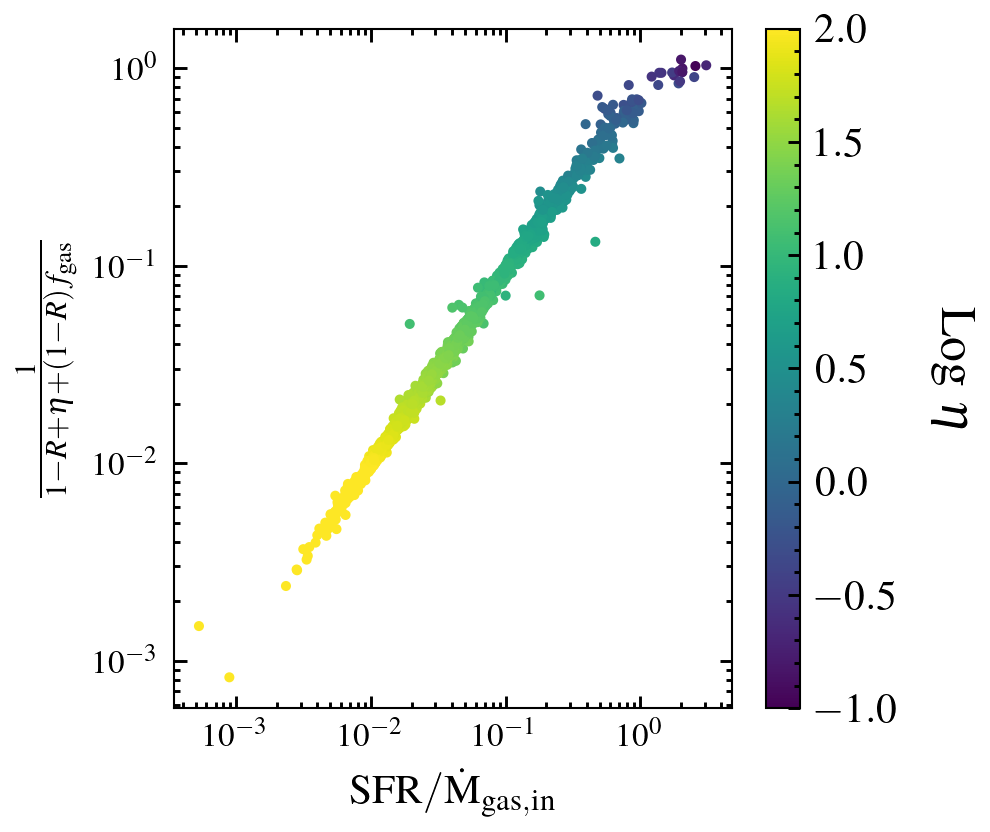}
    \caption{Correlation between the two definitions or $r$ given in Eq.~\ref{eq:r_factor_eta} and Eq.~\ref{eq:r_gas_fraction} color-coded by the value of $\eta$. The two definitions follow the 1:1 line with a small scatter, apart at large values of $r$ where $\eta$ saturates at values $\ll 1$. }
    \label{fig:r_factor}
\end{figure}

\section{Gas fraction as a function of redshift}\label{appendix:gas_fractions}

\begin{figure}
    \centering
    \includegraphics[width=1.0\linewidth]{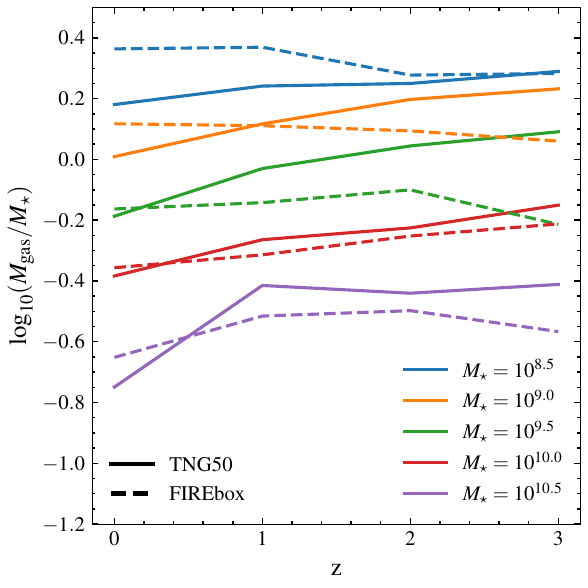}
    \caption{Correlation between the two definitions or $r$ given in Eq.~\ref{eq:r_factor_eta} and Eq.~\ref{eq:r_gas_fraction} color-coded by the value of $\eta$. The two definitions follow the 1:1 line with a small scatter, apart at large values of $r$ where $\eta$ saturates at values $\ll 1$. }
    \label{fig:fgas_z}
\end{figure}

In Fig~\ref{fig:fgas_z} we show the redshift evolution of the gas fraction ($f_{\rm gas}\equiv M_{\rm gas} / M_{\star}$) for two state-of-the-art cosmological simulations: FIREbox and TNG50. For this plot, we compute both the stellar mass and the gas mass within a radial aperture of $0.1\times R_{\rm vir}$.  

At high stellar masses, the two simulations predict similar results, with a predicted redshift evolution of $\sim 0.3$ dex from $z=3$ to $z=0$. at $M_{\star}< 10^{10} M_{\odot}$, TNG50 predicts a stronger evolution than FIREbox, as the latter simulation predicts no evolution in this stellar mass range. However, even in TNG50, the gas fraction in these lower stellar mass bins does not evolve more than  $0.3$ dex. This evolution is not strong enough to explain the redshift evolution of the MZR.

\section{Physical properties as a function of redshift}\label{Appendix:z_fit}

\begin{figure}
    \centering
    \includegraphics[width=0.9\linewidth]{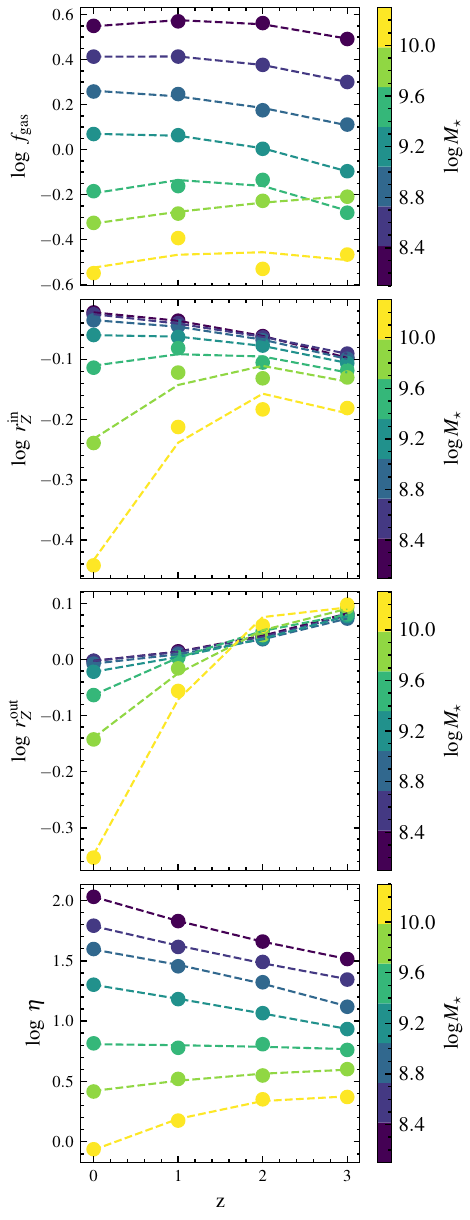}
    \caption{Redshift evolution of inflow metallicity ($r_Z^{\rm in}=Z_{\rm inflow}/Z_{\rm ISM}$), outflow metallicity ($r_Z^{\rm out}=Z_{\rm outflow}/Z_{\rm ISM}$), gas fraction ($f_{\rm gas}$), and mass loading factor($\eta$) as a function of redshift for different stellar mass bins. Results of a fit in the form $\log \xi_i = \alpha + \beta \times z + \gamma \times z^2$ are shown as dashed colored lines.}
    \label{fig:z_fit}
\end{figure}

\begin{figure}
    \centering
    \includegraphics[width=0.8\linewidth]{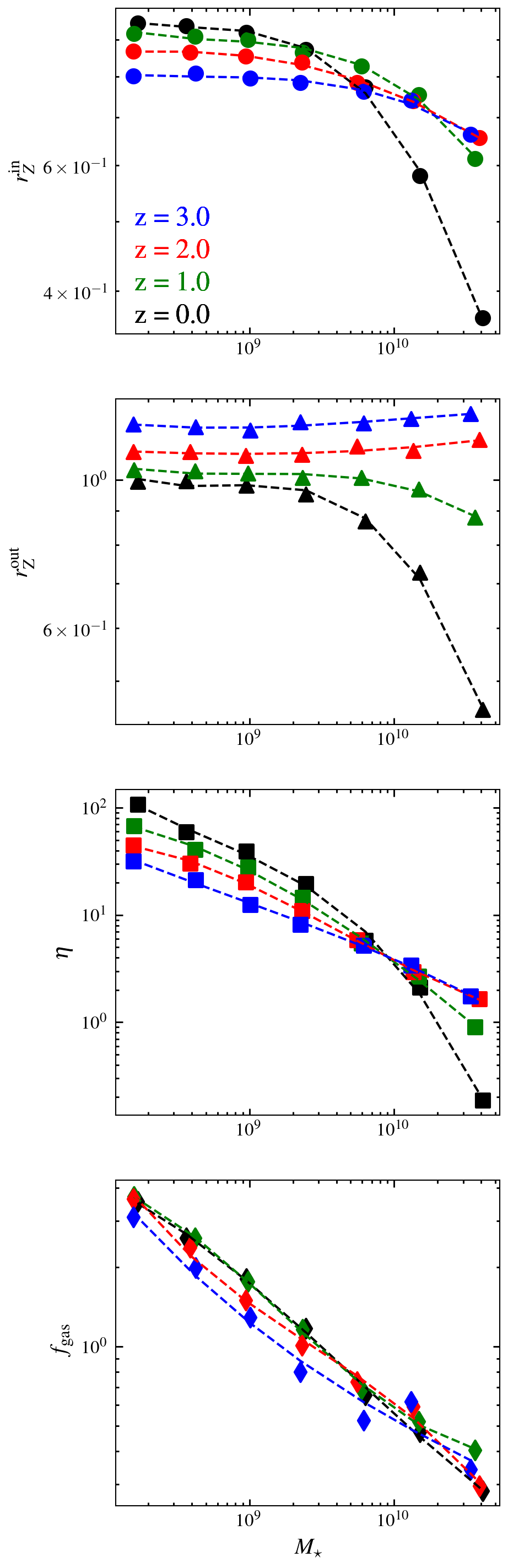}
    \caption{Mass dependence of inflow metallicity ($r_Z^{\rm in}$), outflow metallicity ($r_Z^{\rm out}$), gas fraction ($f_{\rm gas}$), and mass loading factor($\eta$) at fixed redshift ($z=0$, black, $z=1$, green, $z=2$, red, and $z=3$, blue). Each point represents the median in the specific mass bin. Dashed lines show the fit to a polynomial of the third order.}
    \label{fig:z_mass_fit}
\end{figure}

In order to derive the terms in Eq.~\ref{eq:analytical_espression}, an expression for all the $\partial \xi_i / \partial \log (1 + z)$ and $\partial \xi_i / \partial M_{\star}$ are needed. To derive the $\partial \xi_i / \partial \log (1 + z)$ terms, we compute the median values of $\xi_i$ in different mass bins and at $z=0, 1, 2, 3$. Then, we fit the data with a second-order polynomial with the form $\log \xi_i = \alpha + \beta \times z + \gamma \times z^2$. The results of the fitting procedure are shown in Fig.~\ref{fig:z_fit} and listed in Tab.~\ref{tab:xi_sm_fits}. Similarly, to derive the $\partial \xi_i / \partial M_{\star}$ terms, we fit the median values of $\xi_i$ as a function of stellar mass at $z=0, 1, 2, 3$. The data are then fitted with a polynomial of 3$^{\rm rd}$ order. The results of the fit are shown in Fig.~\ref{fig:z_mass_fit} and listed in Tab.~\ref{tab:xi_z_fits}.

\begin{table}
    \centering
    \caption{Results of the second order polynomial fit for the four $\xi_i$ parameters as a function of redshift in different stellar mass bins. Fits are performed in the form $\log \xi_i= \alpha + \beta\times z + \gamma\times z^2$. }
    \begin{tabular}{c|c|c|c|c}
        $\xi_i$ & $\alpha$ & $\beta$ & $\gamma$ & $\log M_{\star}$ \\
        \hline
        \multirow{7}{*}{$r_Z^{\rm in}$} &  -0.022 & -0.008 & -0.005  & 8.0-8.4 \\
         & -0.025 & -0.012 & -0.003 & 8.4-8.8 \\
         & -0.035 & -0.006 & -0.005 & 8.8-9.2 \\ 
         & -0.060 & 0.004  & -0.006 & 9.2-9.6 \\
         & -0.11  & 0.030  & -0.011 & 9.6-10 \\
         & -0.23  & 0.118  & -0.029 & 10-10.4 \\
         & -0.43  & 0.251  & -0.057 & 10.4-10.8 \\
         \hline
         \multirow{7}{*}{$r_Z^{\rm out}$} &  -0.002 & 0.010 & 0.006 & 8.0-8.4 \\
         & -0.002 & 0.011 & 0.005 & 8.4-8.8 \\
         & -0.007 & 0.011 & 0.005 & 8.8-9.2 \\
         & -0.022 & 0.023 & 0.003 & 9.2-9.6 \\
         & -0.063 & 0.076 & -0.009 & 9.6-10 \\
         & -0.13  & 0.13  & -0.019 & 10-10.4 \\
         & -0.35  & 0.34  & -0.065 & 10.4-10.8 \\
         \hline
         \multirow{7}{*}{$f_{\rm gas}$} & 0.55 & 0.05 & -0.023 & 8.0-8.4 \\
         & 0.41  & 0.020 & -0.020 & 8.4-8.8 \\
         & 0.26  & -0.012 & -0.013 & 8.8-9.2 \\
         & 0.070 & 0.015 & -0.024 & 9.2-9.6 \\
         & -0.19 & 0.100 & -0.04 & 9.6-10 \\ 
         & -0.33 & 0.056 & -0.006  & 10-10.4 \\
         & -0.52 & 0.080 & -0.023  & 10.4-10.8 \\
         \hline 
         \multirow{7}{*}{$\eta$} & 2.03 & -0.21 & 0.014 & 8.0-8.4 \\
         & 1.79 & -0.17 & 0.008 & 8.4-8.8 \\
         & 1.59 & -0.11 & -0.015 & 8.8-9.2 \\
         & 1.30 & -0.11 & -0.003 & 9.2-9.6 \\
         & 0.81 & -0.005 & -0.003 & 9.6-10  \\
         & 0.42 & 0.097  & -0.013 & 10-10.4 \\
         & -0.07 & 0.31  & -0.54 & 10.4-10.8 \\
         \hline     
    \end{tabular}
    \label{tab:xi_sm_fits}
\end{table}

\begin{table}
    \centering
    \label{tab:xi_z_fits}
    \caption{Results of the polynomial fit for the four $\xi_i$ parameters as a function of stellar mass at $z=0,1,2,3$. Fits are performed in the form $\log \xi_i= \alpha + \beta\times \log M_{\star} + \gamma \times (\log M_{\star})^2 + \delta \times (\log M_{\star})^3$.}
    \begin{tabular}{c|c|c|c|c|c}
        $\xi_i$ & $\alpha$ & $\beta$ & $\gamma$ & $\delta$ & $z$ \\
        \hline
        \multirow{4}{*}{$r_Z^{\rm in}$} & 31.403 & -10.937 & 1.269 & -0.0491 & 0 \\
        & 18.084 & -6.165 & 0.700 & -0.0265 & 1 \\
        & 1.055 & -0.538 & 0.081 & -0.0038 & 2 \\
        & 7.501 & -2.626 & 0.302 & -0.0116 & 3 \\
        \hline
        \multirow{4}{*}{$r_Z^{\rm out}$} & 42.374 & -14.402 & 1.630 & -0.0615 & 0 \\
        & 12.598 & -4.200 & 0.467 & -0.0173 & 1 \\
        & 0.089 & 0.042 & -0.011 & 0.0007 & 2 \\
        & 3.021 & -0.898 & 0.090 & -0.0029 & 3 \\
        \hline
        \multirow{4}{*}{$f_{\rm gas}$} & -22.840 & 8.0249 & -0.8772 & 0.0301 & 0 \\
        & -51.1722 & 17.6820 & -1.9698 & 0.0711 & 1 \\
        & 60.4485  & -18.4954 & 1.9215 & -0.0678 & 2 \\
        & 20.3909  & -5.1184 & 0.4347  & -0.0129 & 3 \\
        \hline
        \multirow{4}{*}{$\eta$} & 139.7640 & -46.3196 & 5.2431 & -0.2003 & 0 \\
        & -17.8958 & 6.0951  & -0.5646 & 0.0140 & 1 \\
        & -61.9648 & 20.9689 & -2.2477 & 0.0776 & 2 \\
        & 22.1581  & -6.1093 & 0.6329  & -0.0238 & 3 \\
        \hline
    \end{tabular}
\end{table}

\bsp	
\label{lastpage}
\end{document}